\documentclass[10pt, aps,pra,twocolumn,showpacs,superscriptaddress]{revtex4-2}

\usepackage[colorlinks=true,linkcolor=blue,citecolor=blue,urlcolor=magenta]{hyperref}

\usepackage{amsfonts,mathrsfs,amsmath,amsthm,amssymb}
\usepackage{epsfig}
\usepackage{bm}
\usepackage{array,multirow}
\usepackage{booktabs}
\usepackage{graphicx}
\usepackage{upgreek}
\usepackage[normalem]{ulem}
\usepackage{nicematrix}
\usepackage{pifont}

\definecolor{myorange}{rgb}{1.0, 0.55, 0.0}

\begin{document}  
\title {Boosting the performance of a Lipkin-Meshkov-Glick quantum battery via symmetry-breaking quenches or a single-mode bosonic charger}

\author{Le Bin Ho} 
\thanks{Electronic address: binho@fris.tohoku.ac.jp}
\affiliation{Frontier Research Institute 
for Interdisciplinary Sciences, 
Tohoku University, Sendai 980-8578, Japan}
\affiliation{Department of Applied Physics, 
Graduate School of Engineering, 
Tohoku University, 
Sendai 980-8579, Japan}
\affiliation{Theoretical Sciences Visiting Program, Okinawa Institute of Science and
Technology Graduate University, Onna, 904-0495, Japan}

\author{Duc Tuan Hoang} 
\affiliation{Quantum Systems Unit, OIST Graduate University, Onna, Okinawa 904-0495, Japan}

\author{Tran Duong Anh-Tai} 
\affiliation{Quantum Systems Unit, OIST Graduate University, Onna, Okinawa 904-0495, Japan}
\affiliation{Homer L. Dodge Department of Physics and Astronomy, The University of Oklahoma, 440 W. Brooks Street, Norman, Oklahoma 73019, USA}
\affiliation{Center for Quantum Research and Technology, The University of Oklahoma, 440 W. Brooks Street, Norman, Oklahoma 73019, USA}

\author{Thomas Busch} 
\affiliation{Quantum Systems Unit, OIST Graduate University, Onna, Okinawa 904-0495, Japan}

\author{Thom\'{a}s Fogarty} 
\affiliation{Quantum Systems Unit, OIST Graduate University, Onna, Okinawa 904-0495, Japan}

\date{\today}

\begin{abstract}
We explore the operation of quantum batteries in the Lipkin-Meshkov-Glick (LMG) model, when they are charged either through a sudden quench in the magnetic field strength or by coupling them to a single-mode bosonic charger. Through initializing the battery in either the symmetric or broken symmetry phases of the LMG model we analyze how the different spectral properties can affect the performance of both the charging and discharging of the battery. In particular, we show that by quenching the magnetic field strength from the symmetric phase to the broken phase, we can achieve a significant enhancement in stored energy, as well as stable and efficient ergotropy extraction. Similar observations can be made when introducing weak coupling between the battery with the bosonic charger, while the amount of stored work and ergotropy saturate  at strong coupling. These findings emphasize the importance of the magnetic field dynamics or environmental coupling in optimizing charging performance, which could lead to practical applications in quantum energy storage.
\end{abstract}
%
%
\maketitle

\section{Introduction} 
In the epoch of the second quantum revolution, quantum batteries (QBs) are emerging as a promising energy-storage innovation for advancing quantum technologies, particularly for applications necessitating quick charging of quantum devices with high energy-conversion efficiency. In contrast to classical batteries which rely predominantly on chemical compounds, QBs harness quantum properties such as entanglement and coherence to surpass their classical counterpart in various aspects, such as charging and discharging performances, and energy capacity \cite{RevModPhys.96.031001,Binder2019}. 

Over the last couple of years, various QB models have been proposed, each possessing distinct strengths and limitations. Early studies have focused on many-body systems, such as the Dicke model \cite{PhysRevA.100.043833,PhysRevLett.120.117702, PhysRevB.102.245407,PhysRevB.99.205437,Binder_2015,PhysRevB.105.115405}, the Sachdev-Ye-Kitaev model \cite{Rosa2020,PhysRevLett.125.236402}, and spin-chains \cite{PhysRevB.109.235432,PhysRevA.97.022106,PhysRevA.103.033715,PhysRevResearch.4.013172,Barra_2022,PhysRevA.106.032212,PhysRevE.104.024129,PhysRevB.100.115142,PhysRevA.101.032115,Carrega_2020,PhysRevApplied.14.024092,PhysRevA.105.022628,PhysRevA.103.052220,Caravelli2021energystorage,PhysRevB.104.245418,PhysRevE.105.054115,PhysRevE.105.044125,PhysRevE.108.064106,PhysRevResearch.6.013038,PRXQuantum.5.030319}. These models aim to enhance charging efficiency through collective quantum effects \cite{PhysRevA.100.043833,PhysRevLett.120.117702,PhysRevE.104.024129}, entanglement \cite{PhysRevE.101.062114,PhysRevLett.120.117702,PhysRevLett.118.150601,Binder_2015,PhysRevE.102.052109}, non-Hermitian terms \cite{PhysRevA.109.042207}, noise-assisted fast charging \cite{Shastri2025,PhysRevA.104.032207}, and optimal control \cite{Rodriguez_2024,PhysRevLett.133.243602,PhysRevA.107.032218,10762673}. Recent QB designs, such as those based on cavity quantum electrodynamics (cavity QED), utilize strong coupling between qubits and cavity modes \cite{Beleno_2024}, incorporate memory effects \cite{https://doi.org/10.1002/qute.202400115}, and repeated interactions \cite{PhysRevResearch.5.013155} to improve energy storage and charging efficiency. However, achieving a balance between efficiency, robustness, and scalability poses significant challenges. For instance, quantum decoherence restricts expanding the size of these batteries, while limited degrees of freedom in the charger prevent full saturation, which decreases the energy transfer efficiency. Theoretical studies on QBs have thus far examined numerous mechanisms to enhance performance. These include nonreciprocal energy transfer to reduce losses \cite{PhysRevLett.132.210402}, the derivation of upper limits for energy storage \cite{PhysRevLett.132.210402,PhysRevResearch.2.023113}, the use of (non)completely positive trace-preserving maps \cite{PhysRevLett.132.240401}, and the exploration of causal order effects \cite{PhysRevLett.131.240401}. Some studies have integrated topological properties to increase the robustness against noise \cite{PRXQuantum.5.030319,lu2024topologicalquantumbatteries}. Furthermore, experimental investigations of QBs have been conducted using nuclear magnetic resonance (NMR) systems with star topology and photonic platforms \cite{PhysRevLett.131.240401,YANG2024102300,PhysRevB.108.L180301}. Although these approaches demonstrate distinct advantages compared to classical systems, practical challenges, such as complex initialization and control, remain significant limitations. 

In this work, we introduce and analyze a QB based on the Lipkin-Meshkov-Glick (LMG) model, which is a many-body spin system originally studied in nuclear physics \cite{LIPKIN1965188,MESHKOV1965199,GLICK1965211}. From the theoretical perspective, the LMG model has gained attention in the field of quantum thermodynamics due to its collective long-range spin interactions \cite{PhysRevA.78.052101} and second-order phase transition \cite{PhysRevA.80.012318,PhysRevLett.49.478,PhysRevB.71.224420,PhysRevB.94.184403,PhysRevLett.99.050402,PhysRevE.78.021106,PhysRevB.28.3955}. Furthermore, it can be solved exactly using the algebraic Bethe ansatz \cite{MORITA2006337, PAN19991} or by mapping angular momentum operators to \(N\) Schwinger bosons in the weak interaction limit \cite{ORTIZ2005421}. In the thermodynamic limit, the Holstein-Primakoff transformation provides an exact solution \cite{PhysRevLett.93.237204,PhysRev.58.1098,PhysRevE.78.021106}. It is interesting to remark that near the critical point this system exhibits enhanced sensitivity, a characteristic that has been widely explored in quantum sensing, metrology and speed limits \cite{PhysRevE.78.032103,PhysRevE.78.051126,PhysRevA.90.022111,Garbe_2022,montenegro2024reviewquantummetrologysensing, OC_QSL}. Some preliminary works on the LMG model have shown that quantum entanglement can boost charging power \cite{PhysRevResearch.2.023113}, while shortcuts to adiabaticity further improve charging speed and overall performance \cite{dou2022chargingadvantageslipkinmeshkovglickquantum}. More importantly, the LMG model has already been experimentally implemented in trapped ions \cite{Islam2011,Richerme2014,PhysRevLett.119.080501,Zhang2017} and ultra-cold quantum gases \cite{PhysRevLett.105.204101,Hoang2016,PhysRevLett.116.155301}. Overall, these features therefore make the LMG model a compelling candidate for energy storage and energy extraction, with the potential for improved charging efficiency and robustness against decoherence. 

In this study, we systematically investigate the charging protocols of QB by suddenly quenching it across the critical point, or by connecting it to a single-mode bosonic charger.
By quenching the magnetic field from the symmetric phase to the broken symmetry phase and introducing coupling with a  bosonic charger, we observed a significant improvement in stored energy, along with stable and efficient ergotropy extraction. Our results highlight the magnetic field dynamics and environmental coupling as key factors in optimizing charging performance for QBs.

The manuscript is structured as follows. Section \ref{sec2} outlines the theoretical framework and setup for the LMG-based QB. We presents the main findings of our study in Section \ref{sec3}. Finally, the conclusions encompassing a discussion of the implications and potential applications of this approach in quantum energy storage technologies are drawn in Section \ref{sec5}. To simplify notations, we shall refer to the battery Hamiltonian as $H_\mathrm{B}$ and the charging Hamiltonian as $H_\mathrm{C}$ throughout this work. 

\section{Theoretical framework}\label{sec2}
\subsection{The LMG model}
To model the QB, we consider an ensemble of \(N\) spin-$\dfrac{1}{2}$ particles with all-to-all interactions characterized by the LMG Hamiltonian~\cite{LIPKIN1965188,MESHKOV1965199,GLICK1965211}
\begin{align}\label{eq:H_LMG}
H_{\rm LMG} = -\dfrac{\lambda}{N} \sum_{i<j}^{N} \Bigg( \sigma_x^i \sigma_x^j + \gamma \sigma_y^i \sigma_y^j \Bigg) - h \sum_{i=1}^{N} \sigma_z^i,
\end{align}
where \(\sigma_{\alpha}\) are the Pauli matrices with \(\alpha = x, y, z\). Here, \(\lambda\) and \(h\) denote the spin-spin coupling strength, and the strength of the external magnetic field, respectively. The anisotropy parameter \(\gamma\) can range from 0 (fully anisotropic) to 1 (fully isotropic). It has been proven that the LMG model exhibits a second-order quantum phase transition (QPT) driven by the interplay between the spin-spin interactions and the external magnetic field \cite{PhysRevA.80.012318,PhysRevLett.49.478,PhysRevB.71.224420, PhysRevB.94.184403, PhysRevLett.99.050402,PhysRevE.78.021106, PhysRevB.28.3955}. In the thermodynamic limit (\(N \to \infty\)), when \(h > \lambda\), the spins align with the external magnetic field, resulting in spontaneous magnetization. Meanwhile, for \(h < \lambda\), the interaction energy dominates, leading to a two-fold degenerate energy spectrum. The symmetry breaking occurs at the critical point \( h = \lambda \), separating the symmetric phase (\(h > \lambda\)) from the symmetry-broken phase (\(h < \lambda\)). Throughout this study, we will focus on the ferromagnetic case (\( \lambda > 0 \)) and set \(\lambda = 1\) for simplicity. 

For the analysis of the LMG-based QB, one needs the complete energy spectrum and the corresponding eigenstates of the Hamiltonian Eq.~\eqref{eq:H_LMG}. For numerical purposes, it is convenient to rewrite the LMG Hamiltonian given by Eq.~\eqref{eq:H_LMG} in terms of the collective spin operators, \(J_\alpha = \dfrac{1}{2}\sum\limits_i\sigma_\alpha^{(i)}\), which can then be recast as
\begin{align}\label{eq:H_LMG_J}
H_{\rm LMG} = -\dfrac{2}{N}\bigg( J_x^2 + \gamma J_y^2 \bigg) -2 h J_z.
\end{align}
Due to the permutation invariance of the collective spin system \cite{PhysRevA.78.052101}, i.e., the total angular momentum $\mathbf{J}$ is conserved, $[\mathbf{J}, H_{\rm LMG}] = 0$, we only focus on the maximum angular momentum \( j_{\rm max} = N/2 \), resulting in a Hilbert space of dimension \( d = N+1 \). The matrix elements of the Hamiltonian Eq.~\eqref{eq:H_LMG_J} are known analytically in the Dicke basis \( |N/2, m\rangle \equiv |m\rangle \) with \( -N/2 \le m \le N/2 \) and the matrix can then be numerically diagonalized. Further details on the derivation of Eq.~\eqref{eq:H_LMG_J} and the numerical methodology are presented in Appendi.~\ref{appA}. 

\subsection{Charging protocols}
\begin{figure}[t]
\includegraphics[width=\columnwidth]{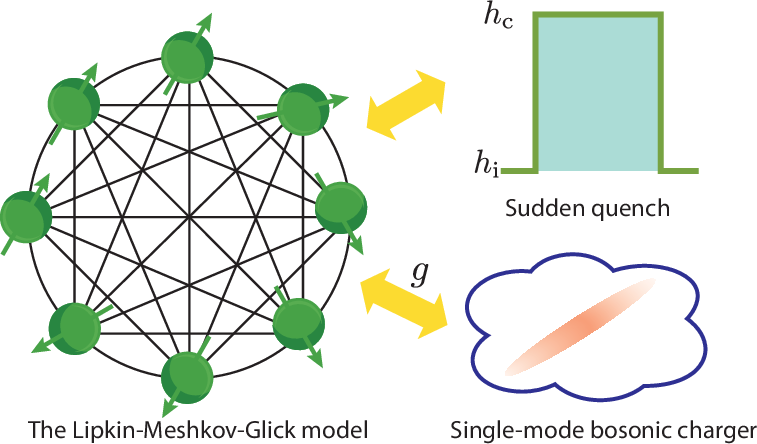}
\caption{An LMG-based QB featuring full-range interactions between spin-1/2 particles. The charging process is employed through (1) sudden quenching of the external magnetic field, (2) coupling to a single-mode bosonic charger.}
\label{fig1}
\end{figure}

In this section, we outline two different charging strategies for an LMG-based QB which are schematically illustrated in Fig.~\ref{fig1}. 

\subsubsection{Charging by sudden quench of the magnetic field}
Initially, the QB is assumed to be in the ground state \( |\varepsilon_0\rangle \) of the LMG Hamiltonian, $H_{\rm B} = H_{\rm LMG}(h_{\rm i})$, with magnetic field strength $h_{\rm i}$ and with initial energy \(\varepsilon_0\). To charge the QB, the magnetic field is abruptly changed from the initial value $h_{\rm i}$ to a value $h_{\rm c} \neq h_{\rm i}$ at time $t = 0$, leading to the creation of out-of-equilibrium dynamics that excites the system to higher energy levels, thereby injecting energy into the QB. As the post-quench dynamics of the QB is governed by the charging Hamiltonian $H_{\rm C} = H_{\rm LMG}(h_{\rm c})$, the charging performance can be optimized by adjusting the magnetic field strength $h_{\rm c}$ along with the initial magnetic field strength $h_{\rm i}$. Our aim is to examine the role of the different phases of the LMG model on both charging and discharging the battery. Specifically, whether quenching within the same phase, or between the phases, can result in more powerful and stable battery performance.

\subsubsection{Charging by coupling to a bosonic charger}
\label{sec:bath}

In the second strategy, the QB is charged by connecting it to a single-mode bosonic charger. Such a system corresponds to a single mode prepared in a Fock state, which leads to a coherent energy exchange between the
battery and a finite-dimensional system. In contrast to a genuine environment with many degrees of freedom, this setup describes coherent energy exchange between the battery and a finite-dimensional system.
Specifically, we consider the QB being prepared in the ground state of the LMG Hamiltonian, $H_\mathrm{B} = - H_\mathrm{LMG}(h_\mathrm{i})$, while the charging Hamiltonian is given by 
\begin{align}\label{eq:HC}
    H_{\rm C} = H_\mathrm{B} + H_{\rm charger} + H_{\rm coupling}.
\end{align}
The coupling between the LMG-based QB and a bosonic charger is modeled as 
\begin{align}\label{eq:Hcoupling}
    H_{\rm coupling} = g (J_+ a + J_- a^\dagger),
\end{align}
where $g$ denotes the coupling strength between the QB and the charger. The charger Hamiltonian is described by $H_{\rm charger}=\omega a^\dagger a$ with $a^\dagger$ and $a$ being the bosonic creation and annihilation operators corresponding to a photon with frequency $\omega$. Here $J_\pm = J_x \pm iJ_y$ are the collective spin raising and lowering operators, respectively (see Appendix~\ref{appA} for details). To fully charge the QB via the coupling in Eq.~\eqref{eq:Hcoupling}, the charger must supply sufficient energy to populate the battery’s maximum excited manifold (i.e., its full capacity). This is achieved by preparing the charger in the Fock state \(|n\rangle\) with \(n \ge N\) \cite{RevModPhys.96.031001,PhysRevResearch.2.023113,PhysRevLett.120.117702}. Note that while many protocols assume this condition in practice, it is not a universal thermodynamic constraint that the charger energy needs to strictly exceed the QB energy for all coupling models \cite{PhysRevLett.120.117702}, so throughout this work we set $n = N$.

\subsection{Quantities of interest}
At time $t$, the energy of the QB is defined as the expectation value of \( H_{\rm B} \) with respect to the time-evolving state $|\psi(t)\rangle =\exp(-iH_\mathrm{C}t)|\psi_0\rangle$
\begin{align}\label{eq:Et}
    E(t) = \langle \psi(t) | H_{\rm B} | \psi(t) \rangle,
\end{align}
where $|\psi_0\rangle$ is the initial state, i.e., $|\psi_0\rangle = |\varepsilon_0\rangle$ for the quenching case and $|\psi_0\rangle = |\varepsilon_0\rangle\otimes |n\rangle$ for the coupling case. 

The total work stored in the entire QB is given by
\begin{align}\label{eq:Work}
    \mathcal{W}(t) = E(t) - \varepsilon_0. 
\end{align}
The power can therefore be defined as
\begin{align}\label{eq:Power}
    \mathcal{P}(t) = \dfrac{\mathcal{W}(t)}{t}.
\end{align}

It is worth noting that only a certain amount of the total work is necessarily accessible, especially when extracting work from only a subset of battery cells. When focusing on \( 0< M \leq N \) cells, this partial discharge might be due to limitations in accessing the full state of the system or intentionally leaving part of the battery charged. In such cases, energy can remain ``locked" in correlations between the \( M \) active cells and the \( N - M \) inactive ones, reducing the total amount of extractable work. Consider an \( M \)-cell subsystem with the Hamiltonian \(H_{\rm B}^M = \sum\limits_{i=1}^{M} \zeta_i|\zeta_i\rangle\langle \zeta_i|\), with \(\zeta_i<\zeta_{i+1}\) and its quantum state at time $t$ is given by
\begin{align}\label{eq:rhoM}
    \rho^M(t) = {\rm Tr}_{N-M}\big[|{\psi(t)}\rangle\langle{\psi(t)}|\big] = \sum\limits_{j=1}^{M} \eta_j(t) |\eta_j(t)\rangle\langle \eta_j(t)|,
\end{align}
with $\eta_j(t) \ge \eta_{j+1}(t)$ being the eigenvalues sorted in descending order. The ergotropy $\mathcal{E}(t)$ of the $M$-cell subsystem with respect to $H^M_{\rm B}$ quantifies the maximum extractable work extracted from $\rho^M(t)$ via a unitary transformation
\begin{align}\label{eq:ep}
\mathcal{E}(t) = \text{Tr}\Big[\rho^M(t) H^M_{\rm B}\Big] - \text{Tr}\Big[\sigma^M(t) H_{\rm B}^M\Big],
\end{align}
where $\sigma^M(t)$ is a passive state associated with $\rho^M(t)$, defined as 
\begin{align}\label{eq:sig}
\sigma^M(t) = \sum_{i=1}^{M} \eta_i(t) |{\zeta_i}\rangle\langle{\zeta_i}|\,. 
\end{align}
This definition ensures that the passive state \(\sigma^M(t)\) possesses a minimum energy, known as the passive energy, thereby preventing any further work from being extracted from \(\rho^M(t)\) via unitary operations~\cite{RevModPhys.96.031001}. The ergotropy \(\mathcal{E}(t)\) therefore can be written in the form
\begin{align}\label{eq:ergoM}
    \mathcal{E}(t) = \sum\limits_{i=1}^M \left[p_i(t) - \eta_i(t)\right] \zeta_i,
\end{align}
where 
\begin{equation}
    \label{eq:pop_passive}
    p_i(t) = \sum\limits_{j=1}^{M}\eta_j(t)|\langle\eta_j(t)|\zeta_i\rangle|^2,
\end{equation}
is the population of \(\rho^M(t)\) 
in the eigenbasis of $H^M_{\rm B}$. It is apparent that the differences between the populations $p_i(t)$ and $\eta_i(t)$ are required for finite work extraction from $M$ active cells. The $M$-cell stored work \(\mathcal{W}^M\) is therefore given as
\begin{align}\label{eq:workM}
    \mathcal{W}^M (t) = {\rm Tr}\big[\rho^M(t)H_{\rm B}^M\big]-{\rm Tr}\big[\rho^M(0)H_{\rm B}^M\big],
\end{align}
where $\rho^M(0)$ is the initial state of the $M$-cell subsystem. 
In general, the ergotropy $\mathcal{E}(t)$ and the $M$-cell stored work $\mathcal{W}^M(t)$ satisfy the inequality $\mathcal{E}(t) \le \mathcal{W}^M(t)$. The equality $\mathcal{W}^M(t) = \mathcal{E}(t)$ holds only when the stored energy is fully extractable, for instance when $\rho^M(t)$ is a pure state and $\rho^M(0)$ is passive; in this case, the passive state coincides with the ground state, leading to $\mathcal{E}(t) = {\rm Tr}[\rho^M(t)H_{\rm B}^M] - E_0 = \mathcal{W}^M(t)$.

It is important to remark that we will trace out $N-M$ cells to construct \(\rho^M(t)\) when examining the charging strategy through magnetic field quenching in section \ref{sec:quenching}. Meanwhile we only trace out the charger and investigate the entire $N$-cell battery in the second strategy via the coupling to a bosonic charger in section \ref{sec:coupling}. 
This distinction arises because the full battery state remains pure in the quench protocol, and tracing out $N-M$ spins is required to generate a mixed state and capture the effects of correlations between spin partitions on energy extraction. In contrast, for the bosonic charger protocol, tracing out the charger already yields a mixed battery state due to battery–charger correlations, and therefore no further spin partitioning is necessary.

\section{Results and discussion}
\label{sec3}
\subsection{Charging by quenching the external magnetic field}
\label{sec:quenching}
In this section, we will discuss the charging protocol for the LMG QB by quenching the magnetic field, which is experimentally feasible. We remark that in the following, our analysis focuses on the anisotropic case (\(\gamma = 0\)) since the system can be excited easily during the charging process and thus store energy more efficiently. In Appendix~\ref{appB}, we demonstrate that the LMG model in the isotropic case (\(\gamma = 1\)) is not a good candidate for realizing QBs as the excitations are mostly suppressed in the broken phase ($h<\lambda$). For the analysis below, we consider \(N = 100\) spin-$1/2$ particles with other parameters specified in each case.  

\subsubsection{The excitation energy spectrum}
First, let us analyze the excitation energy spectrum of the LMG model in the anisotropic case, which will provide the understanding of how much energy can be stored in the battery and its dynamical properties. Let us denote \(\{|\varepsilon_l\rangle\}\) and \(\{\varepsilon_l\}\) the eigenstates and eigenvalues of \( H_{\rm B} \); and \(\{|\mu_k\rangle\}\) and \(\{\mu_k\}\) are the eigenstates and eigenvalues of \( H_{\rm C} \).

\begin{figure}[t]
\includegraphics[width = \columnwidth]{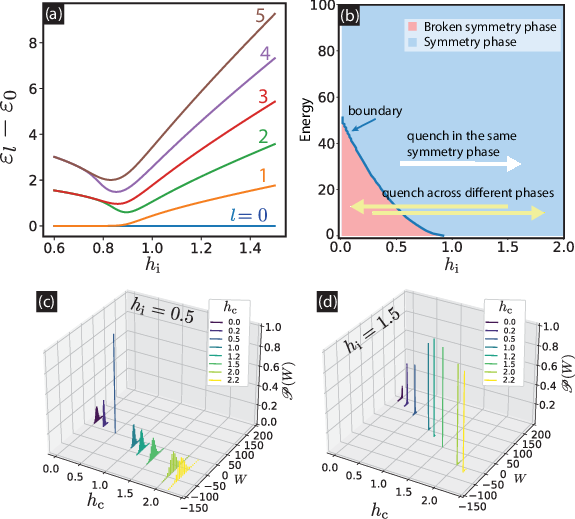}
\caption{(a) Energy gap between the ground state and the first five excited states as a function of the magnetic field \( h_{\rm i} \) for \( N = 100 \) particles in the anisotropic case \( \gamma = 0 \). (b) Phase diagram showing the boundary between the symmetric and broken phase, including cases of quenches within the same phase and across different phases. (c,d) Work probability distribution \(\mathscr{P}(W)\) with different $h_{\rm c}$ for (c) \( h_{\rm i} = 0.5 \) and (d) \( h_{\rm i} = 1.5 \).}
\label{fig2}
\end{figure}

In Fig.~\ref{fig2}(a), we show the energy gap between the ground state and low-lying excited states, \(\varepsilon_l - \varepsilon_0\). When \(h_{\rm i} \gtrsim 0.9\), the system is in the symmetric phase with nondegenerate energy levels. For \(h_{\rm i} < 0.9\), the system transitions to the broken phase, where certain gaps close, creating two-fold degeneracies, consistent with Ref.~\cite{PhysRevB.94.184403}. The critical value of \(h_{\rm i}\) varies with the energy \(\epsilon_l\), and the set of these critical values defines the boundary between the symmetric and broken phases. Fig.~\ref{fig2}(b) presents the phase diagram for \(N = 100\) and in the following we will consider both quenches within the same phase and between different phases. 

\subsubsection{Work probability distribution function}
To deepen our understanding of the charging process, we must analyze the distribution of excitations in the eigenbasis of the charging Hamiltonian $H_\mathrm{C}$ after the magnetic field quench. This can be conveniently quantified by the work probability distribution (WPD) \cite{campisi2011colloquium,PhysRevResearch.4.033014}
\begin{align}
    \mathscr{P}(W) = \sum_k |\langle\mu_k|\varepsilon_0\rangle|^2\ \delta\big(W-({\mu_k-\varepsilon_0})\big),
\end{align}
which quantifies the probability of doing some work $W$ on the system and gives additional insight into how energy is injected into the QB during the charging process.

In Fig.~\ref{fig2}(c) we show the WPD for an initial state being in the broken phase \(h_{\rm i} = 0.5\). In this case, the system can be excited into broad distributions of different energy eigenstates whether quenched within the same broken symmetry phase $h_{\rm{c}}<0.9$ or quenched across the critical point into the symmetric phase $h_{\rm{c}}>0.9$ \cite{PhysRevB.94.184403}. The wide spread of the WPD when the initial state is prepared in the broken phase indicates that quenching in either direction  creates ample excitations in the QB state that could be extracted at a later time.  

In stark contrast, when the system is initiated in the symmetric phase, we can see that $\mathscr{P}(W)$ strongly depends on the quench direction as shown in Fig. \ref{fig2}(d) for the specific case \(h_{\rm i} = 1.5\). For quenches within the symmetric phase (\(h_{\rm c} > 0.9\)), the WPD is extremely localized with a single predominant peak which corresponds to the system remaining in the ground state of the post-quench Hamiltonian, i.e. $|\langle \varepsilon_0 | \mu_0 \rangle|^2 \approx 1$, and there are scant excitations to higher energy eigenstates. Therefore, negligible charging transpires for quenches within the symmetric phase which is due to the fact that the spins are well-oriented by the magnetic field, leading to an absence of interactions and no buildup of many-body excitations. However, for quenches to the broken phase (for instance \(h_{\rm c} < 0.9\)), the interaction between spins becomes dominant and the work distribution broadens, signifying an increased energy contribution once \(h_{\rm c} \to 0\) as depicted in Fig.~\ref{fig3} which we will discuss in the next section.

\subsubsection{Stored energy, entropy, and power}
We explore the total work stored \(\mathcal{W}(t)\) in the battery state at the end of the charging process as calculated from Eq.~\eqref{eq:Work}. Explicitly, it is given by
\begin{align}\label{eq:Work_expl}
    \mathcal{W}(t) = \sum_l \Big|\sum_k \exp({-i\mu_k t})\langle\mu_k|\varepsilon_0\rangle\langle\varepsilon_l|\mu_k\rangle\Big|^2(\varepsilon_l - \varepsilon_0).
\end{align}
Furthermore, we will employ the long-time averaged work
\begin{align}
    \langle \mathcal{W} \rangle = \sum_l \sum_k |\langle\mu_k|\varepsilon_0\rangle|^2 |\langle\varepsilon_l|\mu_k\rangle|^2 (\varepsilon_l - \varepsilon_0),
\end{align}
and its variance 
\begin{align}
\text{Var}(\mathcal{W}) &=
\sum_{l, l^\prime} \sum_k |\langle \mu_k | \varepsilon_0 \rangle|^2 |\langle \varepsilon_l | \mu_k \rangle|^2 |\langle \varepsilon_{l^\prime} | \mu_k \rangle|^2 \ \times \nonumber \\ 
&\hspace{1.7cm}(\varepsilon_l - \varepsilon_0)(\varepsilon_{l^\prime} - \varepsilon_0) \nonumber\\
    &- \left[ \sum_l \sum_k |\langle\mu_k|\varepsilon_0\rangle|^2 |\langle\varepsilon_l|\mu_k\rangle|^2 (\varepsilon_l - \varepsilon_0) \right]^2,
\end{align}
to evaluate the effectiveness of the charging protocols (see Appendix~\ref{appC} for the derivation).

\begin{figure*}[t]
\includegraphics[width = 0.8\linewidth]{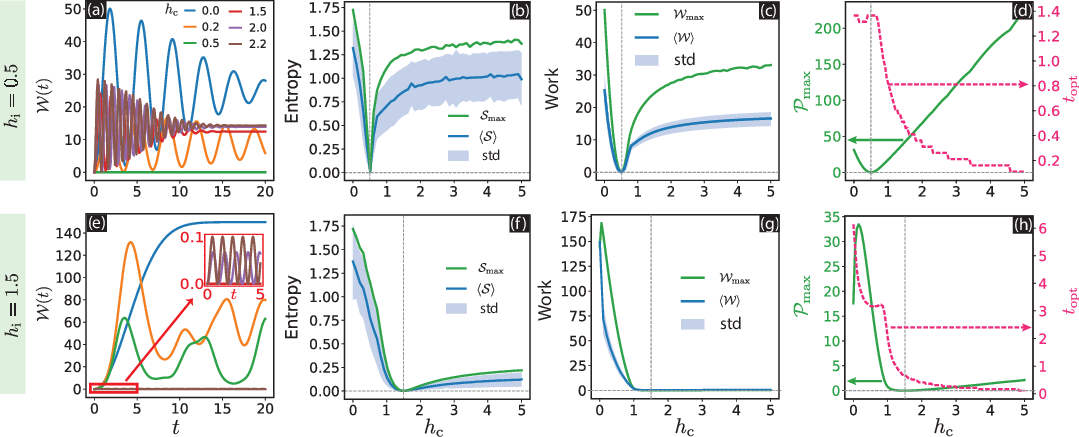}
\caption{Time evolution of the total stored work \(\mathcal{W}(t)\) (a,e). (b,f) Plot of the maximum entropy, \(\mathcal{S}_{\rm max}\), the long-time averaged entropy \(\langle \mathcal{S} \rangle\) and its standard deviation ($\sqrt{\text{Var}(\mathcal{S})}$) as functions of \(h_{\rm c}\). (c,g) Plot of the maximum stored work, \(\mathcal{W}_{\rm max}\), averaged work \(\langle \mathcal{W} \rangle\) and its standard deviation ($\sqrt{\text{Var}(\mathcal{W})}$) as functions of \(h_{\rm c}\). (d,h) Variation of the peak power, \(\mathcal{P}_{\rm max}\), and the optimal charging time, \(t_{\rm opt}\), as functions of \(h_{\rm c}\), illustrating contrasting dynamical behaviors. The first and second rows represent the initial state being in the broken phase (\(h_{\rm i} = 0.5\)) and the symmetric phase (\(h_{\rm i} = 1.5\)), respectively. The inset in panel (e) magnifies the interval \(0 \leq t \leq 5\), highlighting short-time dynamics. }
\label{fig3}
\end{figure*}

\textbf{Broken phase:} We first discuss initializing the QB in the broken phase with \(h_{\rm i} = 0.5\) and then quench to different final values of the magnetic field strength \(h_{\rm c}\). The stored work in the QB is shown in Fig.~\ref{fig3}(a) as a function of the charging time for representative values of \(h_{\rm c}\). Quenching within the same phase (\(h_{\rm c} < h_{\rm i}\)) the average work \(\mathcal{W}(t)\) has large oscillations which slowly decay [blue and orange curves in Fig. \ref{fig3} (a)]. When no quench occurs (\(h_{\rm c} = h_{\rm i} = 0.5\)), there is no energy gain and therefore $\mathcal{W}(t)=0$. In contrast, quenching across the critical point (\(h_{\rm c} = 1.5, 2.0, 2.2\)) results in faster oscillations and a quicker saturation of the stored work. 

This underlying physics can be understood from the excitation spectrum in Fig.~\ref{fig2}(a). In the broken phase, the closing of energy gaps and near-degeneracies allow multiple low-energy excitations to participate in the dynamics. As seen from Eq.~\eqref{eq:Work_expl}, this leads to many contributing frequencies, enhancing quantum fluctuations and producing large-amplitude, slowly decaying oscillations of $\mathcal{W}(t)$ for quenches within the same phase. In contrast, quenching across the critical point drives the system into a regime with larger gaps, reducing the number of excitation channels and resulting in faster oscillations and quicker saturation of the stored work.

This behavior is also reflected in the Shannon entropy shown in Fig.~\ref{fig3}(b), which allows us to quantify how spread the energy population is in the battery Hamiltonian. It is defined as
\begin{align}\label{eq:entropy}
\mathcal{S}(t) = -\sum_i p_i(t)\,\ln p_i(t),
\end{align}
where the populations $p_i$ are given in Eq.~\eqref{eq:pop_passive} for $M = N$. When no quench occurs ($h_{\rm c} = h_{\rm i} = 0.5$), the maximum entropy \(\mathcal{S}_{\rm max}\), and the average entropy over time \(\langle\mathcal{S}\rangle\) remain zero, consistent with the absence of energy gain. In contrast, when the magnetic field is quenched, both \(\mathcal{S}_{\rm max}\) and \(\langle\mathcal{S}\rangle\) are finite indicating that the system explores a broader set of excited states which is consistent with the enhanced work storage observed in Fig.~\ref{fig3}(a) and the work probability distribution shown in Fig.~\ref{fig2}~(c). Thus, the entropy dynamics reinforce the work analysis: quenching drives the system deeper into the excited spectrum, enabling more efficient charging regardless of the direction of the quench, $h_{\rm c}<h_{\rm i}$ or $h_{\rm c}>h_{\rm i}$, as shown in the vertical dashed line in Fig.~\ref{fig3}(b).

To clarify the effectiveness of the charging of the LMG battery over the entire operation range of $h_{\rm c}$, we show the maximum work \(\mathcal{W}_{\rm max}\) and the average work \(\langle \mathcal{W} \rangle\) in Fig~\ref{fig3}(c). 
Quenching to smaller magnetic field strengths $h_{\rm c}<h_{\rm i}$ results in larger values of maximum work and its time average when compared to quenching to larger magnetic field strengths $h_{\rm c}>h_{\rm i}$. After the critical point at $h_{\rm c}\approx 0.9$ the time-averaged work only slowly increases with increasing magnetic field strength, appearing to saturate when driven deep into the symmetric phase. We also note that the standard deviation of the stored work, indicated by the light blue shaded region, possesses larger values in this regime, indicating increased fluctuations of the energy that is stored in the battery.

Finally, we analyze the maximum power $\mathcal{P}_{\rm max}$ and the optimal charging time $t_{\rm opt}$, as shown in Fig.~\ref{fig3}(d). The results show that $\mathcal{P}_{\rm max}$ is small when the quench is performed within the same broken phase but increases significantly when the system is quenched into the symmetric phase. At the same time, $t_{\rm opt}$ decreases monotonically as $h_{\rm c}$ increases. This is due to quenches within the broken phase creating slow oscillations, resulting in a large $t_{\rm opt}$ and a small $\mathcal{P}_{\rm max}$. In contrast, quenching into the symmetric phase induces faster oscillations, producing a smaller $t_{\rm opt}$ and a much larger $\mathcal{P}_{\rm max}$.

\textbf{Symmetric phase:} Next, we discuss preparing the battery in the symmetric phase with \(h_{\rm i} = 1.5\) and again quenching to the new magnetic field strength $h_{\rm c}$. The stored work \(\mathcal{W}(t)\) is shown in Fig.~\ref{fig3}(e), and immediately it is clear that the behavior of the battery is drastically different. When quenching within the symmetric phase, \(\mathcal{W}(t)\) oscillates without decaying, with vanishingly small values as shown in the inset. This behavior is consistent with the spectrum in Fig.~\ref{fig2}(a). In the symmetric phase, the spectrum is non-degenerate with larger energy gaps, limiting the number of accessible excitations and leading to small, regular oscillations of $\mathcal{W}(t)$.
This is further supported by the analysis of the work probability distribution in Fig.~\ref{fig2}(d) and the entropy shown in Fig.~\ref{fig3}(f), indicating that quenches within the symmetric phase generate only a few excitations and thus result in minimal energy storage in the QB state.

However, significant features appear when quenching across the phase boundary from the symmetric to the broken phase. The further the system is driven into the broken phase the more work is stored in the QB. For finite magnetic field strengths \(h_{\rm c} \neq 0\), the maximum of the work is quickly reached before displaying complex oscillations. However, in the limit \(h_{\rm c} = 0\), \(\mathcal{W}(t)\) increases at a slower rate but eventually saturates to a large and constant value. 
Here, quenching into the broken phase introduces smaller gaps and degeneracies, allowing more excitation channels to contribute and resulting in enhanced energy spreading and improved work storage, as seen in Fig.~\ref{fig2}(a).
We note that in this case the maximum and average work are equivalent as seen in Fig.~\ref{fig3}(g), as the variance of the stored work effectively vanishes, displaying a significant charging advantage. Again, this analysis is consistent with the work probability distribution shown in Fig.~\ref{fig2}(d) and the entropy as shown in Fig.~\ref{fig3}(f), as the energy is distributed over a larger region of the Hilbert space when quenched deeper into the broken phase, leading to significantly larger energy storage when compared to $h_{\rm i}=0.5$. Lastly, we note that while more work can be stored in this QB, it requires longer charging times which reduces the power, as shown in Fig.~\ref{fig3}(h).

\subsubsection{Analytical understanding of enhanced charging across the critical point}

To provide a simple analytical explanation of the enhanced charging observed in our numerical results, we treat the quench as a perturbation. Since only the magnetic field is changed, the battery Hamiltonian can be written as
\begin{align}
H_{\rm B} = H_{\rm C} + \delta h\, V, 
\;
\delta h = h_i - h_c,
\;
V = 2J_z.
\end{align}
Let $\{|\mu_k\rangle, \mu_k\}$ denote the eigenstates and eigenvalues of $H_{\rm C}$, and let $|\varepsilon_0\rangle$ be the ground state of $H_{\rm B}$.
Expanding $|\varepsilon_0\rangle$ in the eigenbasis of $H_{\rm C}$ using first-order perturbation theory yields
\begin{align}
|\varepsilon_0\rangle
\approx
|\mu_0\rangle
+
\delta h \sum_{k>0}
\frac{\langle \mu_k | V | \mu_0 \rangle}{\mu_0 - \mu_k}
|\mu_k\rangle.
\end{align}
The excitation probabilities are therefore
\begin{align}
p_k \equiv |\langle \mu_k | \varepsilon_0 \rangle|^2
\approx
\delta h^2
\frac{|\langle \mu_k | V | \mu_0 \rangle|^2}{(\mu_k - \mu_0)^2},
\qquad k>0,
\end{align}
and $p_0 = 1 - \sum_{k>0} p_k$. Defining the quench energy gaps $\Delta_k = \mu_k - \mu_0$, we obtain
\begin{align}
p_k \propto \frac{1}{\Delta_k^2}.
\end{align}

This result shows that the excitation probability depends directly on the energy gaps of \(H_{\rm C}\). When the system is quenched from the symmetric phase into the symmetry-broken phase, the low-energy gaps become small and nearly degenerate. As a result, the excitation probabilities \(p_k\) increase, and the initial state spreads over many low-energy states.
In contrast, if the quench stays within the symmetric phase, the spectrum remains well separated and non-degenerate. The energy gaps are large, so \(p_k \ll 1\). In this case, the system stays close to the ground state, and only a small amount of energy is stored.
This explains why quenches across the critical point lead to much higher stored energy: The small energy gaps in the symmetry-broken phase make it easier for the system to absorb energy.
This is related to the same basic mechanism discussed in LMG thermodynamic-cycle studies, namely that critical spectral structure and degeneracy enhance thermodynamic performance \cite{PhysRevE.96.022143, PhysRevLett.134.027101}, but here it appears in a non-equilibrium quench battery setting rather than an equilibrium heat-engine cycle.

\subsubsection{Extracted energy}
To explore the role of correlations on the amount of energy that can be extracted from the QB, we calculate the ergotropy \(\mathcal{E}\) given by Eq.~\eqref{eq:ergoM} for a subsystem of $M\leq N$ spins. Quantum correlations between these $M$ spins and the remaining $N-M$ spins of the total system will lead to a reduction of the ergotropy with respect to the amount of work  \(\mathcal{W}^M\) stored in this subsystem given by Eq.~\eqref{eq:workM}. We can therefore define the ratio between the ergotropy and the work,  \(\mathcal{E}/\mathcal{W}^M\), as a quantifier of the efficiency of the energy extraction process, saturating to unity when no correlations are present.  

\begin{figure}[t]
\includegraphics[width = \columnwidth]{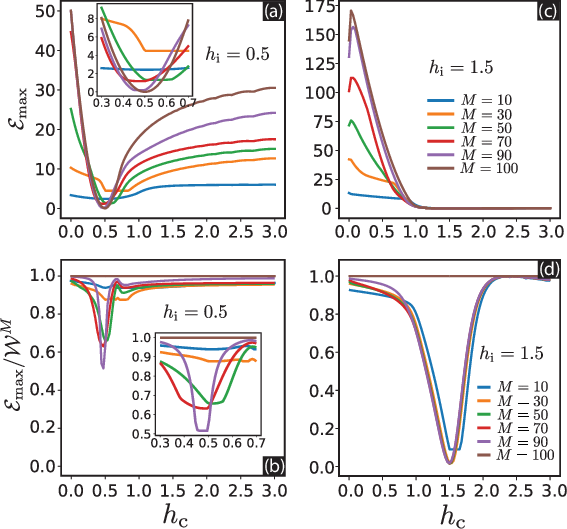}
\caption{The maximum ergotropy \(\mathcal{E}_{\rm max}\) and the efficiency ratio \(\mathcal{E}_{\rm max}/\mathcal{W}^M\) as a function of $h_{\rm c}$ for various values of \(M\) (see the legends). Panels (a-b) show the results for \(h_{\rm i} = 0.5\), whereas panels (c) and (d) illustrate those for \(h_{\rm i} = 1.5\). Insets: zoom in from $h_{\rm c} = 0.3$ to 0.7. Numerically, we add a small offset \(10^{-10}\) to \(\mathcal{W}^M\) to avoid undefined values when \(h_{\rm c} = h_{\rm i}\).
}
\label{fig4}
\end{figure}

\begin{figure*}[t]
\includegraphics[width = 0.8\linewidth]{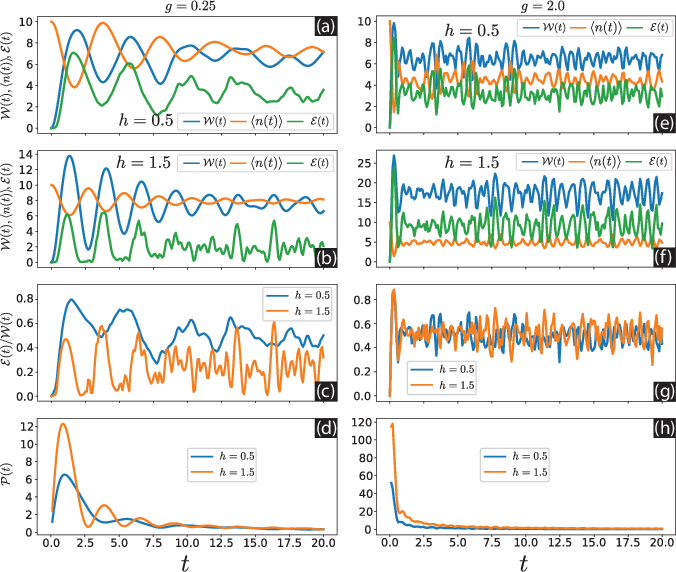}
\caption{Time evolution of the stored energy $\mathcal{W}(t)$, the average photon number $\langle n(t) \rangle$, the ergotropy $\mathcal{E}(t)$, the ratio $\mathcal{E}(t)/\mathcal{W}(t)$,
and power $\mathcal{P}(t)$, for two coupling strengths: $g = 0.25$ (left column) and $g = 2.0$ (right column).}
\label{fig5}
\end{figure*}

The full time-dependent results of the ergotropy and work are shown in App.~\ref{appE} for the different quenching strategies. From these data, we extract the maximum total ergotropy \(\mathcal{E}_{\rm max}\) that is reached after the quench and its ratio \(\mathcal{E}_{\rm max}/\mathcal{W}^M\) with the corresponding amount of work. In Fig.~\ref{fig4}(a,b), we show these quantities as a function of the charger magnetic field strength $h_{\rm c}$ starting in the broken phase (\(h_{\rm i} = 0.5\)). For $M = N$, \(\mathcal{E}_{\rm max} = \mathcal{W}_{\rm max}^M\). As \(M\) decreases, \(\mathcal{E}_{\rm max}\) similarly decreases as expected, as less energy can be extracted from a smaller numbers of cells. Near the initial magnetic field strength, $h_{\rm c}\approx h_{\rm i}$, the efficiency ratio \(\mathcal{E}_{\rm max}/\mathcal{W}^M\) is diminished as shown in Fig.~\ref{fig4}(b). However, it increases when quenching to either side of \(h_{\rm i}\) reaching values above 0.9, highlighting the robustness of the LMG battery against detrimental correlation effects.

For \(h_{\rm i} = 1.5\), similar results are depicted in Figs.~\ref{fig4}(c,d). Quenching from the symmetric phase into the broken phase offers a clear advantage in terms of the ergotropy, while again the ratio takes large values \(\mathcal{E}_{\rm max}/\mathcal{W}^M> 0.9\) for $h_{\rm c}<1$. However, when quenching within the symmetric phase, $h_{\rm c}>1$, although the ratio \(\mathcal{E}_{\rm max}/\mathcal{W}^M\) can take large values, the ergotropy itself is always very small, as shown in Fig.~\ref{fig4}(c), indicating that this scenario is less favorable for any meaningful energy storage and extraction. 

Finally, we also note that in both cases at \( h_{\rm c} = h_{\rm i} \), \(\mathcal{E}_{\rm max}\) remains nonzero for \( M < N \) due to the inherent ergotropy in the $M$ cell subsystem arising from the interactions with the remaining spins $N-M$ cells that are present in the charged state. As the passive state does not include this interaction energy it is therefore extracted from the subsystem even when no charging quench is applied, and contributes a minor increase in ergotropy for $h_{\rm c}\neq h_{\rm i}$ as also noted in Ref.~\cite{PhysRevB.107.075116}.

\subsection{Charging by coupling to a bosonic charger}
\label{sec:coupling}
In this section, we analyze the charging efficiency of the QB through coupling to a bosonic charger, as described in \ref{sec:bath}. Here, we fix $N = M = 10$ spins. The charger is initialized as a Fock state $|n\rangle$, with $n$ photons (here we consider \( n = N = 10 \)) . The battery parameters are fixed with \( \lambda = 1 \), and full anisotropy \( \gamma = 0 \). We consider two values of the external magnetic field, $h = h_{\rm i} = 0.5$ (broken phase) and $h = h_{\rm i} = 1.5$ (symmetric phase) in the following analysis so as to only examine the role the charger plays in the charging process. For numerical calculations the Fock space is truncated to 50 states which is sufficiently large to avoid finite-size effects for initial photon numbers up to 10.

We first focus on the regime of small battery-charger coupling $g=0.25$, and show the stored work $\mathcal{W}(t)$, average photon number $\langle n(t)\rangle$ and the ergotropy $\mathcal{E}(t)$, for $h=0.5$ and $h=1.5$ in Fig.~\ref{fig5}(a) and Fig.~\ref{fig5}(b) respectively. For both magnetic field strengths, the stored work and average photon number exhibit damped oscillations that are out of phase: initially, $\mathcal{W}(t)$ increases from zero while $\langle n(t)\rangle$ decreases from its initial value of $n=10$. This behavior reflects the exchange of energy between the charger and the battery which is induced by the sudden interaction quench. 

\begin{figure}[t]
\includegraphics[width = \linewidth]{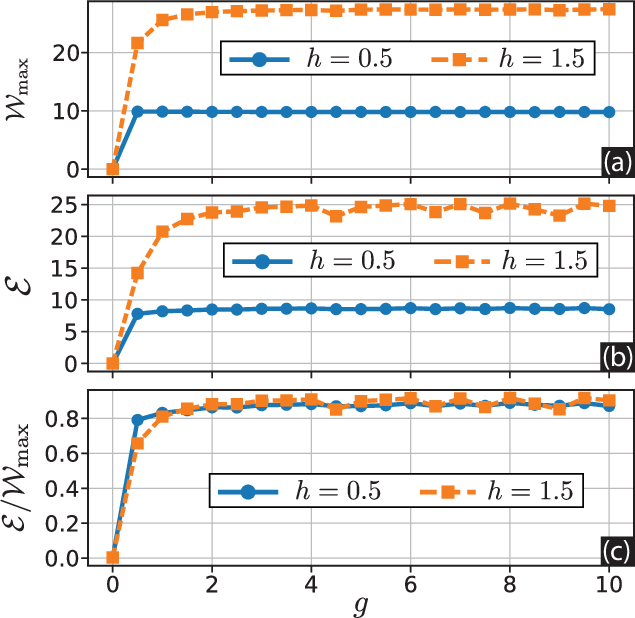}
\caption{(a) The maximum stored work $\mathcal{W}_{\rm max}$, (b) the ergotropy $\mathcal{E}$ at $\mathcal{W}_{\rm max}$, and (c) the ratio $\mathcal{E}/ \mathcal{W}_{\rm max}$ (c) as a function of coupling strength $g$ for $h=0.5$ (blue dotted line) and $h=1.5$ (orange dotted line).}
\label{fig6}
\end{figure}

While qualitatively the charging process is similar for both $h=0.5$ and $h=1.5$, the amount of work that can actually be extracted via the ergotropy is quite different. Indeed, while more work is stored in the symmetric phase battery, the maximum ergotropy is actually comparable to the broken phase battery. Furthermore, the ergotropy can practically vanish in the symmetric phase battery when the stored work takes local minima at $t\approx 2.5$ and $t\approx 5$, while this is not the case for the broken phase battery. This difference in performance and energy transfer efficiency can be succinctly seen in the ratio $\mathcal{E}(t)/\mathcal{W}(t)$ shown in Fig.~\ref{fig5}(c), which shows that on average a larger amount of extractable work can be stored in the broken phase battery, and which crucially is more stable on longer times.   

When the coupling between the battery and charger is large, $g=2$, we again observe clear signatures of energy exchange between the battery and the charger, indicated by the reduction in the average photon number and subsequent increase in work and ergotropy (see Fig.~\ref{fig5}(e,f)). For the broken phase battery, the long time behavior of the stored work and the ergotropy is equivalent to the weak coupling case (see Fig.\ref{fig5}(a)), albeit with faster oscillations caused by the increased battery-charger coupling. However, for the symmetric battery significantly more work is stored and ergotropy extracted, more than double in both cases, than in the same battery at weak coupling (see Fig.\ref{fig5}(b)). Furthermore, the ratio $\mathcal{E}(t)/ \mathcal{W}(t)$ is now equivalent for both broken phase and symmetric phase batteries, as shown in Fig.~\ref{fig5}(g), indicating similar performance in terms of the percentage of energy that can be extracted.

We also examine the charging power $\mathcal{P}(t) = \mathcal{W}(t)/t$, shown in Figs.~\ref{fig5}(d) and \ref{fig5}(h) for $g = 0.25$ and $g = 2.0$, respectively. In both phases, the power exhibits a sharp peak at short times before decaying, indicating that the most efficient energy injection occurs at the early stages of the charging process. Increasing the coupling strength from $g = 0.25$ to $g = 2.0$ leads to a significantly larger peak power and a shorter optimal charging time, consistent with the faster oscillations observed in the stored work. Notably, the symmetric phase ($h = 1.5$) achieves a much higher peak power than its broken phase counterpart ($h = 0.5$) at strong coupling, with the peak power increasing by nearly an order of magnitude. This highlights that strong coupling to the bosonic charger not only enhances the total stored energy but also substantially improves the charging rate, making the symmetric phase battery more favorable when fast and powerful charging is required.

While this implies that stronger coupling leads to more powerful batteries, we find that already at $g=2$ the maximum performance is reached in this system. We can see this in Fig.~\ref{fig6} where we show the the maximum stored work $\mathcal{W}_{\rm max}$, ergotropy $\mathcal{E}$ at $\mathcal{W}_{\rm max}$, and their ratio as functions of the coupling strength $g$. For the broken phase battery, the maximum stored work and ergotropy have essentially saturated to a fixed value at 
$g=0.5$, while for the symmetric battery this occurs around $g=2$.  

Comparing the maximum stored and extracted work of the two batteries we find that both $\mathcal{W}_{\rm max}$ and $\mathcal{E}$ attain significantly higher values for the symmetric battery with $h = 1.5$. This behavior can be understood by examining the energy spectrum of the QB shown in Fig.~\ref{fig2}(a). At $h = 0.5$, the system exhibits a nearly degenerate energy gap $\varepsilon_l - \varepsilon_0$, which limits the energy that can be stored. In contrast, at $h = 1.5$, the energy levels are non-degenerate and the energy gaps increase with $h$, thereby raising the upper bound of storable energy. This observation is consistent with earlier studies \cite{PhysRevResearch.2.023113,PhysRevA.103.052220}, which demonstrated that the form of the energy spectrum plays a crucial role in determining the maximum stored energy in a QB. In our case, the field-dependent energy spectrum sets a natural upper limit for $\mathcal{W}_{\rm max}$ and $\mathcal{E}$. The observed saturation with increasing $g$ indicates that this limit is eventually reached, and stronger coupling cannot further increase the battery's storage capacity.

\begin{figure}[t]
\includegraphics[width = \linewidth]{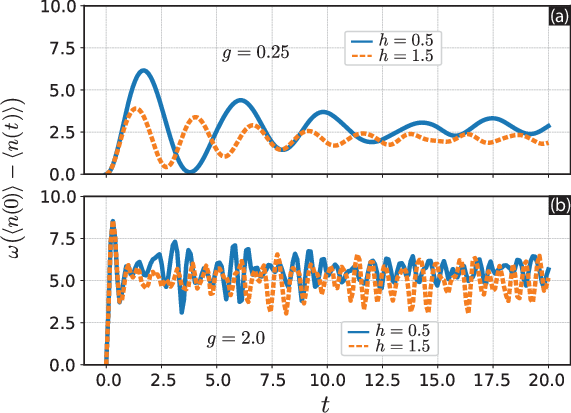}
\caption{Plot of $E_{\rm mode}(t)$ as a function of time for two cases of $g = 0.25$ (a) and $g = 2.0$ (b).}
\label{fig7}
\end{figure}

Since the interaction term in Eq.~\eqref{eq:HC} does not commute with $H_{\rm B}$, switching on the coupling between the QB and the charger injects irreversible work into the system. Consequently, the energy stored in the QB at the end of the charging process originates not only from the charger but also from the external work performed during charging.
To make this distinction explicit, we evaluate the energy extracted from the bosonic mode as 
\begin{equation}
    E_{\mathrm{mode}}(t) = \omega\big(\langle n(0)\rangle - \langle n(t)\rangle\big).
\end{equation}
We show the result of $E_{\mathrm{mode}}(t)$ in Fig.~\ref{fig7}.
As we mentioned above, $\mathcal{W}(t)$ increases while $\langle n(t)\rangle$ decreases. This behavior reflects the exchange of energy between the charger and the battery. 

The broken phase generally exhibits a larger energy exchange between the charger and the battery, particularly in the weak-coupling regime. This suggests that the enhanced energy storage observed in the symmetric phase originates primarily from irreversible work generated during the quench rather than from energy transfer. The lower ergotropy ratio in the symmetric phase further supports this interpretation, indicating that a larger fraction of the stored energy remains locked in charger-battery correlations. At stronger coupling, the energy exchange becomes comparable in both phases, whereas the symmetric phase continues to store more energy, highlighting the dominant contribution of irreversible work.

\begin{figure}[b]
\includegraphics[width = \linewidth]{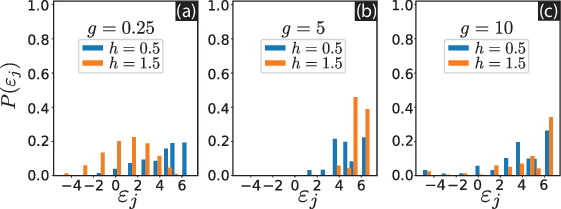}
\caption{Occupations $P(\varepsilon_j)$ of the eigenstates of battery Hamiltonian  \(H_{\rm B}\)  as a function of the eigenenergies $\varepsilon_j$ for battery-charger couplings (a) $g=0.25$, (b) $g=5$ and (c) $g=10$.}
\label{fig8}
\end{figure}

To further support our results, we show in Fig.~\ref{fig8} the energy distribution $P(\varepsilon_j) = \langle\zeta(\varepsilon_j)|\rho_{\rm B}|\zeta(\varepsilon_j)\rangle$ as a function of eigenvalues $\varepsilon_j$, where $\{|\zeta(\varepsilon_j)\rangle\}$ are eigenstates of $H_B$ and $\rho_{\rm B}$ is the battery state after tracing out the charger. The shown results correspond to the same time that \(\mathcal{W}_{\rm max}\) is reached.

Once the coupling \(g\) is turned on, energy begins to flow between the charger and the battery, causing the work to increase and the population to spread into higher energy levels with \(\varepsilon_j>\varepsilon_0\).
In particular, the broken phase battery already occupies higher energy eigenstates for $g=0.25$, while the symmetric phase battery is normally distributed in the centre of the spectrum (see Fig.~\ref{fig8}(a)). As the coupling becomes stronger, the population gradually shifts upward through the energy ladder and eventually approaches the maximum level, \(\varepsilon_{j_{\text{max}}}\), as in Fig.~\ref{fig8}(b). At $g=10$, both cases capacity saturates as levels close to \(\varepsilon_{j_{\text{max}}}\) are occupied as seen in Fig.~\ref{fig8}(c). 
Complete population inversion of either battery is not reached due to irreversible excitations that are created during the coupling process, along with non-Markovian effects from strong coupling to the charger \cite{zhao2025stabilizingergotropyspinchainquantum, Kamin_2020,e24060820,PhysRevLett.129.130602}. This results in lower energy eigenstates of the battery being occupied throughout the charging process leading to a reduction in extractable work as seen in the ergotropy in Fig.~\ref{fig6}(c). 

\section{Conclusion} \label{sec5}
In this paper, we have studied the performance of the LMG-based QB, which is charged by either quenching a magnetic field or coupling to a bosonic charger. We have shown that quenching a symmetric phase QB across the critical point results in the significant enhancement in the stored work, along with stable and efficient ergotropy extraction. If we charge the QB by strongly coupling to the bosonic charger, the symmetric phase QB shows a better performance compared to its broken symmetry counterpart due to its non-degenerated energy levels, while still maintaining the same work extraction ratio. These results highlight the magnetic field dynamics and environmental coupling as key factors in optimizing charging performance.

The study emphasizes that precise control over both the parameters for magnetic field quenching and the strength of the bosonic charger coupling is essential for achieving maximum energy storage efficiency. The LMG model's ability to maintain high ergotropy under these conditions underscores its potential as a robust framework for quantum energy storage.

In addition to demonstrating the advantages of this approach, our findings also pave the way for future investigations. Further research could focus on developing advanced optimization protocols to refine quenching strategies and coupling dynamics. Exploring other types of environmental couplings or extending this approach to multi-qubit quantum batteries could provide deeper insights and enhance the practical applicability of quantum energy storage technologies.

\begin{acknowledgments}
This research was supported by the SHINKA grant at the Okinawa Institute of Science and Technology (OIST) and was also conducted while visiting the OIST through the Theoretical Sciences Visiting Program (TSVP). T.B., T.F., and T.D.A.-T. thank support from JST COI-NEXT Grant No. JPMJPF2221. T.F. is also grateful to JSPS KAKENHI Grant No. JP23K03290. T.D.A.-T. acknowledges support through a Dodge Postdoctoral Researcher Fellowship at the University of Oklahoma.
\end{acknowledgments}

\section*{Data availability}
The data that support the findings of this article are openly
available \cite{Ho2026}.

\section*{Conflict of interest}
The authors declare that there are no conflicts of interest associated with this study.

\appendix

\setcounter{equation}{0}
\renewcommand{\theequation}{A.\arabic{equation}}
\section{The spectral decomposition of the LMG Hamiltonian}
\label{appA}
To numerically obtain the complete spectrum of the Lipkin-Meshkov-Glick (LMG) model described by the Hamiltonian
\begin{align}\label{eq:app_LMG}
  H_\mathrm{LMG} = -\frac{\lambda}{N} \sum_{i<j}^N \left( \sigma_x^i \sigma_x^j + \gamma \sigma_y^i \sigma_y^j \right) - h \sum_{i=1}^N \sigma_z^i,
\end{align}
it is natural to employ the Dicke basis since the matrix elements can be computed analytically within this framework (see Refs.~\cite{PhysRevB.74.104118,PhysRevLett.114.177206,PhysRevB.94.184403} for further details). In doing so, we will use the angular momentum operators $J_\alpha$ defined as
\begin{align}\label{eq:app:J}
  J_\alpha = \frac{1}{2} \sum_{i=1}^N \sigma_\alpha^i,
\end{align}
where $\sigma_\alpha$ with $\alpha=x,y,z$ denotes the Pauli matrices. Straightforwardly, the Zeeman term simply becomes
\begin{align}
    h\sum_{i}\sigma_z^i = 2hJ_z.
\end{align}
Let us then rewrite the two-body term $\sum\limits_{i<j}\sigma_\alpha^i\sigma_\alpha^j$ in Eq.\eqref{eq:app_LMG} in terms of the operator $J_\alpha$ using the expression
\begin{align}\label{app:eqtwo-sum}
\left(\sum_{i=1}^N \sigma_\alpha^i\right)^2 = \sum_{i=1}^N (\sigma_\alpha^i)^2 + 2\sum\limits_{i<j}\sigma_\alpha^i\sigma_\alpha^j.
\end{align}
It is apparent that for the $i$-th particle we have \((\sigma_\alpha^i)^2=\mathbb{I}\), yielding 
\begin{equation}
    \sum\limits_{i=1}^N (\sigma_\alpha^i)^2 = N\mathbb{I},
\end{equation}
with $\mathbb{I}$ being the identity matrix. Hence, we get
\begin{align}\label{app:eq_sum2}
\sum_{i<j}\sigma_\alpha^i\sigma_\alpha^j=\frac{1}{2}\left[\left(\sum_{i=1}^N \sigma_\alpha^i\right)^2 - N\mathbb{I}\right].
\end{align}
Plugging Eq.~\eqref{eq:app:J} into Eq.~\eqref{app:eq_sum2} yields
\begin{align}
\sum_{i<j}\sigma_\alpha^i\sigma_\alpha^j = \frac{1}{2}\left(4J_\alpha^2 - N\mathbb{I}\right) = 2J_\alpha^2 - \frac{N}{2}\mathbb{I}.
\end{align}
Explicitly, we obtain
\begin{equation}
\sum_{i<j}\left(\sigma_x^i\sigma_x^j + \gamma\sigma_y^i\sigma_y^j\right) = 2\left(J_x^2 + \gamma J_y^2\right) - \frac{N}{2}(1+\gamma)\mathbb{I}.
\end{equation}
The Hamiltonian \eqref{eq:app_LMG} is then transformed into
\begin{equation}
\label{eq:app_LMGtrans}
H_\mathrm{LMG} = -\frac{2\lambda}{N}\left(J_x^2 + \gamma J_y^2\right) - 2hJ_z - \frac{\lambda}{2}(1+\gamma)\mathbb{I}.
\end{equation}
It is clear that the term proportional to \(\mathbb{I}\) uniformly shifts the entire spectrum by $-0.5{\lambda}(1+\gamma)$, thereby not influencing neither the eigenstates nor energy differences. Consequently, it is safe and customary to exclude it from Eq.~\eqref{eq:app_LMGtrans} for the sake of simplicity. 

Afterward, we employ the Dicke basis \(|j, m\rangle\), where \(j\) is the total angular momentum quantum number and \(m\) is the magnetic quantum number ranging from \(-j\) to \(j\). For \(N\) spin-1/2 particles, it is noted that the total angular momentum quantum number \(j\) can take values \(0, 1/2, 1, ..., N/2\), however we only focus on the maximum angular momentum quantum number \( j_{\rm max} = N/2 \) due to the permutation invariance of the collective spin system \cite{PhysRevA.78.052101}, ie., the total angular momentum $\mathbf{J}$ is conserved, $[\mathbf{J}, H_{\rm LMG}] = 0$. This leads to a Hilbert space of dimension \(d = N+1 \). For simplicity, we will write \(|j_{\rm max} = N/2, m\rangle\) by \(|m\rangle\) throughout the remaining of this appendix. To further simplify calculations, it is convenient to rewrite the operators $J_x$, and $J_y$ in terms of the collective spin raising and lowering operators (also known as the ladder operators) $J_+$, and $J_-$, which are defined as
\begin{align}\label{eq:appJxJy}
    J_x = \frac{1}{2}\big(J_+ + J_-), \text{ and }
    J_y = \frac{1}{2i}\big(J_+ - J_-).
\end{align}
In the Dicke basis, the eigenvalues and eigenstates of the operators \(J_z\), \(J_+\), and \(J_-\) are respectively given as
\begin{align}\label{eq:app_Jz}
    J_z |m\rangle &= m |m\rangle,\\
    J_+ |m\rangle &= \sqrt{(N/2 - m)(N/2 + m + 1)} |m + 1\rangle,\\
    J_- | m\rangle &= \sqrt{(N/2 + m)(N/2 - m + 1)} |m - 1\rangle,\\
    J_+ J_- | m \rangle &= (N/2 + m)(N/2 - m + 1) | m \rangle,\\
    J_- J_+ | m \rangle &= (N/2 - m)(N/2 + m + 1) | m \rangle,\\
    J_\pm^2 | m \rangle &= \sqrt{(A_\pm+N)(A_\pm-1)}|m\pm2\rangle,
\end{align}
where $A_\pm = N^2/4-m^2\mp2m$. 

With the use of the Dicke basis and the angular momentum operators $J_\alpha$, the matrix representation of the Hamiltonian \eqref{eq:app_LMGtrans} has the banded form as follow: 
\begin{align}
    H_{m^\prime m} & = \langle m^\prime|H|m\rangle \nonumber \\
    & =  \dfrac{\lambda}{2N} (1+\gamma)\sqrt{(A_-+N)(A_- -1)} \delta_{m^\prime,m-2}\nonumber\\
    & + \left[2mh +
    \dfrac{\lambda}{2N} (1+\gamma)(N(0.5N+1) -2m^2) \right]\delta_{m^\prime, m} \nonumber\\
    & + \dfrac{\lambda}{2N} (1+\gamma)\sqrt{(A_++N)(A_+-1)} \delta_{m',m+2}.
\end{align}
We remark that the \((N + 1) \times (N + 1)\) matrix representing the LMG Hamiltonian in the Dicke basis can then be numerically diagonalized to obtain the eigenvalues and eigenstates by using well-known linear algebra packages or computational software. Specifically, we used the function `eigsh' implemented in the SciPy library for the matrix diagonalization in this work.

\setcounter{equation}{0}
\renewcommand{\theequation}{B.\arabic{equation}}
\section{The isotropic LMG-based battery}
\label{appB}
\begin{figure}[t]
\includegraphics[width = \linewidth]{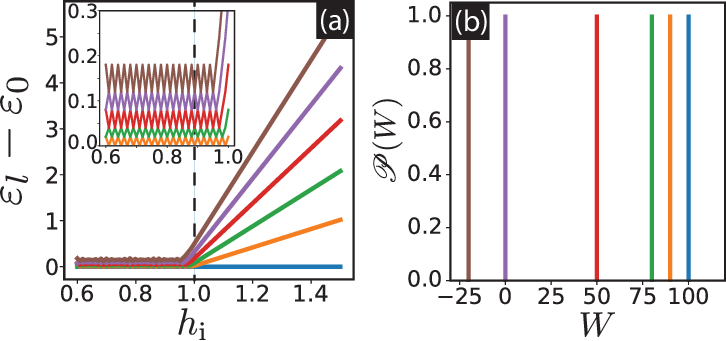}
\caption{(a) Energy gap between the ground state and the first five excited states as a function of \( h_{\rm i} \) for \( N = 100 \) and \( \gamma = 1 \).  (b) Work probability distribution \(\mathscr{P}(W)\) for different $h_{\rm i}$ from right to left: 0.0, 0.1, 0.2, 0.5, 1, 1.2.}
\label{fig9}
\end{figure}

In the following appendix, we consider the LMG-based battery in the isotropic case, \(\gamma = 1\), and demonstrate that this case is not a good candidate for realizing QBs. In the isotropic case, the LMG Hamiltonian is given by
\begin{align}\label{eq:HLMG_iso}
\notag H &= -\dfrac{2}{N} \left( J_x^2 + J_y^2 \right) - 2 h J_z \\
& = -\dfrac{2}{N} \left( \textbf{J}^2 - J_z^2 \right) - 2 h J_z,
\end{align}
where \(\textbf{J}^2 = J_x^2 + J_y^2 + J_z^2\). It can be seen that this Hamiltonian is already diagonal in the Dicke basis \( |m\rangle \), where \( -N/2 \le m \le N/2 \). Consequently, in this situation the eigenstates of the LMG Hamiltonian $|\varepsilon_l\rangle = |N/2-l\rangle$ are exactly the Dicke basis \(|m\rangle \), and the corresponding eigenvalues are
\begin{widetext}
\begin{align}\label{eq:varepl}
\varepsilon_l = 
\begin{cases}
\dfrac{2}{N} \left(\big[\dfrac{Nh}{2}\big]-\dfrac{l}{2} - \dfrac{Nh}{2}\right)^2 + \left(\dfrac{Nh^2}{2} + \dfrac{N}{2} + 1\right) \quad &\text{for } 0 \le h < 1, \\
\dfrac{2}{N} \left(\dfrac{N}{2}-l - \dfrac{Nh}{2}\right)^2 + \left(\dfrac{Nh^2}{2} + \dfrac{N}{2} + 1\right) \quad &\text{for } h > 1,
\end{cases}
\end{align}
\end{widetext}
where $[x]$ denotes the standard rounding function \cite{PhysRevB.71.224420}. As a result, the energy gap $\varepsilon_l - \varepsilon_0$ yields 
\begin{align}\label{eq:delta_varep}
\varepsilon_l - \varepsilon_0 = 
\begin{cases}
	 \dfrac{4l}{N}\Big(\dfrac{Nh}{2}-\big[\dfrac{Nh}{2}\big]+\dfrac{l}{2}\Big) - \dfrac{2}{N}l^2 \quad &\text{for } 0 \le h < 1, \\
     2l\big(h-1\big) - \dfrac{2}{N}l^2 \quad &\text{for } h > 1,
\end{cases}
\end{align}
and closes at $h = (2p+l)/N$ between the two states $|\varepsilon_l\rangle =|m = N/2-l+1\rangle$ and $|\varepsilon_l\rangle =|m = N/2-l\rangle$, for $p \le 0 $ is an integer 
and $l \ge 1$. 
Fig.~\ref{fig9}(a) illustrates the energy gap for the isotropic case. 

We also analyze the work distribution \(\mathscr{P}(W)\), shown in Fig.~\ref{fig9}(b), with \(|\varepsilon_l\rangle = |\mu_l\rangle = |N/2-l\rangle\). In this case, \(\langle\mu_k|\varepsilon_0\rangle \langle\varepsilon_l|\mu_k\rangle = 0\) for all \(l > 0\) and \(k\). As a result, \(\mathscr{P}(W) = 1\) for all \(h_\mathrm{c}\), leading to \(\mathcal{W}(t) = 0\) at all times. This indicates that the energy remains unchanged, meaning that the isotropic LMG model is not a suitable candidate for realizing QBs.

\setcounter{equation}{0}
\renewcommand{\theequation}{C.\arabic{equation}}
\section{Average of the extracted energy in the sudden quench}
\label{appC}
The long-time averaged work \( \langle \mathcal{W} \rangle \) is defined as
\begin{align}
    \label{eq:app:averagework}
    \langle \mathcal{W} \rangle = \lim_{T \to \infty} \frac{1}{T} \int_{0}^{T} \mathcal{W}(t) dt = \sum_l \langle \mathscr{P}_l(t) \rangle (\varepsilon_l - \varepsilon_0).
\end{align}
where the long-time averaged population is given by
\begin{align}
    \langle \mathscr{P}_l(t) \rangle &= \lim_{T \to \infty} \frac{1}{T} \int_0^T \mathscr{P}_l(t) dt \nonumber \\ 
    &= \lim_{T \to \infty} \frac{1}{T} \int_0^T dt \Big[ \nonumber \\
    &\sum_{k,k'} e^{-i (\mu_k - \mu_{k'}) t} \langle \mu_k | \varepsilon_0 \rangle \langle \varepsilon_l | \mu_k \rangle \langle \mu_{k'} | \varepsilon_0 \rangle^* \langle \varepsilon_l | \mu_{k'} \rangle^* \Big].
\end{align}
After some basic algebraic calculations, we obtain
\begin{align}
    \label{eq:app_averagepopulation}
    \langle \mathscr{P}_l(t) \rangle = \sum_{k} |\langle\mu_k|\varepsilon_0\rangle|^2 |\langle\varepsilon_l|\mu_k\rangle|^2,
\end{align}
where the off-diagonal elements with \( k \neq k' \) vanish over the long-time average due to oscillatory integration. Substituting Eq.~\eqref{eq:app_averagepopulation} into Eq.~\eqref{eq:app:averagework} yields
\begin{align}
    \langle \mathcal{W} \rangle = \sum_l \sum_k |\langle\mu_k|\varepsilon_0\rangle|^2 |\langle\varepsilon_l|\mu_k\rangle|^2 (\varepsilon_l - \varepsilon_0).
\end{align}
The variance of the total work \(\mathcal{W}(t)\) over the long-time dynamics is defined as
\begin{align}
    \text{Var}(\mathcal{W}) = \langle \mathcal{W}^2 \rangle - \langle \mathcal{W} \rangle^2,
\end{align}
where
\begin{align}
    \langle \mathcal{W}^2 \rangle & =  \sum_{l, l'} \langle \mathscr{P}_l(t) \mathscr{P}_{l'}(t) \rangle (\varepsilon_l - \varepsilon_0)(\varepsilon_{l'} - \varepsilon_0) \nonumber \\ & = \sum_{l, l'} \sum_k |\langle \mu_k | \varepsilon_0 \rangle|^2 |\langle \varepsilon_l | \mu_k \rangle|^2\ \times\nonumber\\
    &\hspace{2cm}|\langle \varepsilon_{l'} | \mu_k \rangle|^2 (\varepsilon_l - \varepsilon_0)(\varepsilon_{l'} - \varepsilon_0).
\end{align}
Explicitly, the variance of the total work over the long-time dynamics is given as
\begin{align}
\text{Var}(\mathcal{W}) &=
\sum_{l, l^\prime} \sum_k |\langle \mu_k | \varepsilon_0 \rangle|^2 |\langle \varepsilon_l | \mu_k \rangle|^2 |\langle \varepsilon_{l^\prime} | \mu_k \rangle|^2 \ \times \nonumber \\ 
&\hspace{1.7cm}(\varepsilon_l - \varepsilon_0)(\varepsilon_{l^\prime} - \varepsilon_0) \nonumber\\
    &- \left[ \sum_l \sum_k |\langle\mu_k|\varepsilon_0\rangle|^2 |\langle\varepsilon_l|\mu_k\rangle|^2 (\varepsilon_l - \varepsilon_0) \right]^2,
\end{align}

\begin{figure*}[t]
\includegraphics[width = \linewidth]{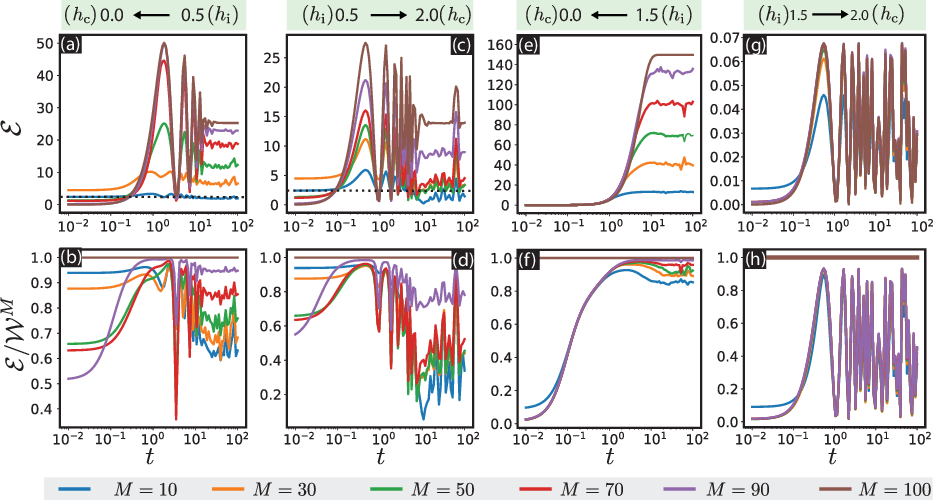}
\caption{The time evolution of the ergotropy \(\mathcal{E}\) and the efficiency ratio \(\mathcal{E}/\mathcal{W}^M\) for different quenching protocols.}
\label{fig10}
\end{figure*}

\setcounter{equation}{0}
\renewcommand{\theequation}{D.\arabic{equation}}

\section{Construction of the reduced density matrix in the Dicke basis}
\label{appD}
We consider a system consisting of $N$ spin-$\dfrac{1}{2}$ particles described in the Dicke basis $|j,m\rangle$, where the total spin quantum number satisfies $j \le N/2$ and $m = -j, \ldots, j$. The full density matrix can be written as
\begin{align}
\rho(t) = \sum_{j_1,m_1} \sum_{j_2,m_2} \rho_{j_1 m_1,\, j_2 m_2}(t)\,
| j_1, m_1 \rangle \langle j_2, m_2 | .
\end{align}

We aim to construct the reduced density matrix $\rho^M(t)$ of a subsystem of $M < N$ spins by tracing out the remaining $N-M$ spins. The reduced subsystem is represented in the Dicke basis $|j', m'\rangle$ with $j' \le M/2$ and $m' = -j', \ldots, j'$.

To perform the partial trace, we use the decomposition of Dicke states into subsystems,
\begin{align}
| j, m \rangle
=
\sum_{j',j''}
\sum_{m',k}
C^{j,m}_{j',m';\, j'',k}
\;
| j', m' \rangle \otimes | j'', k \rangle ,
\end{align}
where $j'$ and $j''$ correspond to the total spins of the $M$-spin subsystem and the $(N-M)$-spin subsystem, respectively, and $C^{j,m}_{j',m';\, j'',k}$ are Clebsch-Gordan coefficients.

Using this decomposition, the reduced density matrix elements are obtained as
\begin{align}
\rho^M_{j'_1 m'_1,\, j'_2 m'_2}(t)
=
\sum_{j_1,j_2}
\sum_{k}
C^{j_1,m_1}_{j'_1,m'_1;\, j''_1,k}
\;
C^{j_2,m_2}_{j'_2,m'_2;\, j''_2,k}
\;
\rho_{j_1 m_1,\, j_2 m_2}(t),
\end{align}
where the indices satisfy $m_1 = m'_1 + k$ and $m_2 = m'_2 + k$, and the sum runs over all allowed values of intermediate quantum numbers.

Finally, the reduced density matrix is normalized as
\begin{align}
\rho^M(t) \rightarrow \frac{\rho^M(t)}{\mathrm{Tr}[\rho^M(t)]},
\end{align}
provided that $\mathrm{Tr}[\rho^M(t)] \neq 0$.

This construction provides the exact partial trace in the Dicke basis, accounting for the full decomposition of the Hilbert space into irreducible spin sectors.

\setcounter{equation}{0}
\renewcommand{\theequation}{E.\arabic{equation}}
\section{Time-dependent of the extracted energy in the sudden quench}
\label{appE}
This section presents the time evolution of the extracted energy via ergotropy \(\mathcal{E}\) and the efficiency ratio \(\mathcal{E}/\mathcal{W}\), where \(\mathcal{E}\) and \(\mathcal{W}\) are defined in Eqs.~(\ref{eq:ergoM}, \ref{eq:workM}) in the main text. The results, shown in Fig.~\ref{fig10}, correspond to different quenching protocols:  
\begin{itemize}
    \item (a,b): Quenching within the broken phase from \(h_{\rm i} = 0.5\) to \(h_{\rm c} = 0.0\).
    \item (c,d): Quenching from the broken phase to the symmetric phase, \(h_{\rm i} = 0.5\) to \(h_{\rm c} = 2.0\).  
    \item (e,f): Quenching from the symmetric phase to the broken phase, \(h_{\rm i} = 1.5\) to \(h_{\rm c} = 0.0\).  
    \item (g,h): Quenching within the symmetric phase from \(h_{\rm i} = 1.5\) to \(h_{\rm c} = 2.0\).  
\end{itemize}

The third case is the most effective charging protocol, while the last case is the least effective. We also emphasize that the relative ergotropy, the extractable energy after the quench compared to the initial state, can become negative at longer times. For example, in Figs.~\ref{fig10}(a, c) for \( M = 10 \), the relative ergotropy turns negative around \( t = 10 \), as indicated by the data going below the dotted black line.

\bibliography{refs}

\end{document}